\documentclass[twocolumn,showpacs,superscriptaddress,preprintnumbers,oneside,pra]{revtex4}
\usepackage{amssymb}
\usepackage{amsfonts}
\usepackage{amsmath}
\usepackage{graphicx}
\usepackage{times}
\usepackage{graphics}
\usepackage{bm}
\usepackage{calc}
\usepackage{fancyhdr}
\usepackage{amsthm}
\usepackage{dcolumn}

\begin{document}

\title{Duality relation and joint measurement in a Mach-Zehnder
Interferometer}
\author{Nai-Le Liu}
\email{nlliu@ustc.edu.cn}
\affiliation{Department of Modern Physics and Hefei National Laboratory for Physical
Sciences at Microscale, University of Science and Technology of China,
Hefei, Anhui 230026, China}
\author{Li Li}
\email{eidos@mail.ustc.edu.cn}
\affiliation{Department of Modern Physics and Hefei National Laboratory for Physical
Sciences at Microscale, University of Science and Technology of China,
Hefei, Anhui 230026, China}
\author{Sixia Yu}
\affiliation{Department of Modern Physics and Hefei National Laboratory for Physical
Sciences at Microscale, University of Science and Technology of China,
Hefei, Anhui 230026, China}
\affiliation{Department of Physics, National University of Singapore, 2 Science Drive 3,
Singapore 117542, Singapore}
\author{Zeng-Bing Chen}
\affiliation{Department of Modern Physics and Hefei National Laboratory for Physical
Sciences at Microscale, University of Science and Technology of China,
Hefei, Anhui 230026, China}

\begin{abstract}
The Mach-Zehnder interferometric setup quantitatively characterizing the
wave-particle duality implements in fact a joint measurement of two unsharp
observables. We present a necessary and sufficient condition for such a pair
of unsharp observables to be jointly measurable. The condition is shown to
be equivalent to a duality inequality, which for the optimal strategy of
extracting the which-path information is more stringent than the
Jaeger-Shimony-Vaidman-Englert inequality.
\end{abstract}

\maketitle

\section{Introduction}

Bohr's principle of complementarity \cite{Bohr}, a statement that a single
quantum system possesses mutually exclusive but equally real properties, is
an essential feature that distinguishes quantum from classical realm. The
best-known manifestation of this principle is the wave-particle duality,
i.e., the fact that a quantum object can at times behave as a wave and at
other times behave as a particle, depending on the circumstances of the
experiment being performed \cite{Scully}. Since the pioneering work of
Wootters and Zurek \cite{WZ}, the conventional qualitative characterization
of the wave-particle duality has acquired its quantitative version. In
particular, in a two-path interferometer such as a Mach-Zehnder
interferometer, there exist two kinds of trade-off relations between the
fringe visibility of the interference pattern and the maximum amount of
which-path information. The first one, known as uncertainty relationship for
preparation, is about the trade-off between the \emph{a priori} fringe
visibility $\mathcal{V}_{0}$ of the interference pattern and the \emph{path
predictability} $\mathcal{P}$ \cite{GY,Mandel}. The measurements of $%
\mathcal{V}_{0}$ and $\mathcal{P}$ can only be carried out by two
incompatible experimental setups since two noncommuting sharp observables
must be measured. The second one is about the trade-off between the fringe
visibility $\mathcal{V}$ and the \emph{path distinguishability} $\mathcal{D}$
obtained simultaneously in a \textit{single} experimental setup equipped
with a which-path detector. Such a setup was first considered independently
by Jaeger \emph{et al.} \cite{Jaeger95} and Englert \emph{et al. }\cite%
{Englert,EnglertBergou}. They established quantitative duality relations
such as%
\begin{equation}
\mathcal{D}^{2}+\frac{1-\mathcal{P}^{2}}{\mathcal{V}_{0}^{2}}\mathcal{V}%
^{2}\leq 1,  \label{JSVE}
\end{equation}%
which we refer to as the Jaeger-Shimony-Vaidman-Englert inequality.

A second manifestation of quantum complementarity is the impossibility of
jointly measuring some pairs of (sharp or unsharp) observables \cite%
{unsharp,jm,jm1,Busch,BS}. The Jaeger-Shimony-Vaidman-Englert setup
implements in fact a joint measurement of two unsharp observables, i.e., a
nonideal joint measurement of two noncommuting sharp observables. A
fundamental question naturally arises: does the condition under which such
two unsharp observables are jointly measurable dictate a duality relation?
In this paper we answer this question in the positive. In Sec. \ref%
{sec:unsharp}, we will give explicit expressions of two unsharp observables
jointly measured in Englert's setup \cite{Englert}. Then, in Sec. \ref%
{sec:condition}, we will derive a necessary and sufficient condition for
such a pair of unsharp observables to be jointly measurable. In Sec. \ref%
{sec:duality} we will show that the condition is \emph{equivalent} to a
duality inequality characterizing the trade-off between the fringe
visibility and the path distinguishability. One will see that the duality
inequality is more stringent than the Jaeger-Shimony-Vaidman-Englert
inequality for the optimal strategy of extracting the which-path
information. We conclude with some discussions in Sec. \ref{sec:conclusion}.

\section{Two unsharp observables jointly measured in Englert's setup}

\label{sec:unsharp}

A standard Mach-Zehnder interferometer as considered by Englert \cite%
{Englert} can be described by a two-dimensional Hilbert space spanned by two
orthonormal states $|0\rangle $ and $|1\rangle $ representing two distinct
paths (see Fig. 1). A generic state of the quanton prior to entering the
interferometer can be represented by a density operator $\rho $ on this
two-dimensional Hilbert space. After passing a beam splitter which, without
loss of generality, is described by the Hadamard transformation $H=(\sigma
_{x}+\sigma _{z})/\sqrt{2}$, the quanton undergoes a phase shifter $\Phi
=e^{i\phi \sigma _{z}/2}$, and then the two beams are combined on another
beam splitter which is also described by the Hadamard transformation.

\begin{figure}[tbp]
\includegraphics[width=70mm]{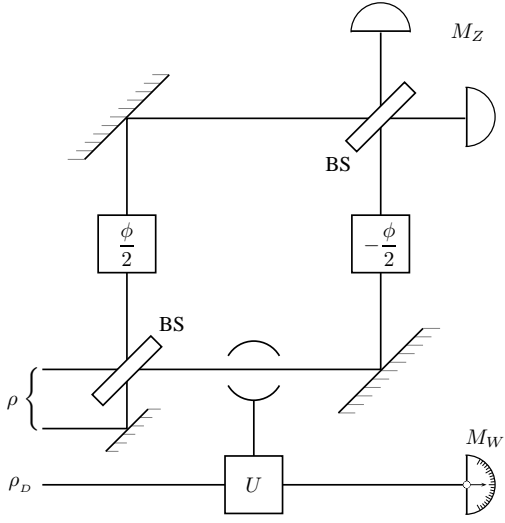}
\caption{The schematic sketch of a Mach-Zehnder interferometer with a
which-path detector. The quanton undergoes a projective measurement $M_Z$ of
$\protect\sigma_z$ after it has passed through the second beam splitter, and
the detector undergoes a projective measurement $M_W$ of $\hat{W}$.}
\end{figure}

Immediately follows from the positivity of the density matrix a quantitative
duality relation $\mathcal{P}^{2}+\mathcal{V}_{0}^{2}\leq 1$ between the
\textit{a priori} fringe visibility $\mathcal{V}_{0}=2\left\vert \langle
+|\rho |-\rangle \right\vert $ and the predictability $\mathcal{P}%
=\left\vert w_{+}-w_{-}\right\vert $ with $w_{\pm }=\langle \pm |\rho |\pm
\rangle $ being the probabilities for the quanton taking the two paths
respectively, where $|\pm \rangle =(|0\rangle \pm |1\rangle )/\sqrt{2}$ are
two eigenvectors of $\sigma _{x}.$ To test this duality relation, two
projective measurements must be made: $\sigma _{x}$ for the predictability $%
\mathcal{P}=|\langle \sigma _{x}\rangle _{\rho }|$ and
\begin{equation}
\sigma _{\phi }=(H\Phi H)^{\dagger }\sigma _{z}(H\Phi H)=\sigma _{z}\cos
\phi -\sigma _{y}\sin \phi
\end{equation}%
for the \textit{a priori} fringe visibility
\begin{equation}
\mathcal{V}_{0}=\max_{\phi }\langle \sigma _{\phi }\rangle _{\rho },
\label{v0}
\end{equation}%
the maximum being attained when $\phi $ is set to be the phase factor $\phi
_{0}$ of $\langle -|\rho |+\rangle $. Note that $\sigma _{x}$ and $\sigma
_{\phi _{0}}$ are a pair of noncommuting sharp observables, whose
measurements cannot be fulfilled in a single experimental setup.

To simultaneously obtain the which-path information and the interference
pattern, a detector is coupled with the quanton by a controlled unitary
transformation $U_{QD}=|0\rangle \langle 0|\otimes I_{D}+|1\rangle \langle
1|\otimes U$ after the quanton passes through the first beam splitter, where
$I_{D}$ and $U$ are respectively the identity operator and a unitary
operator on the Hilbert space of the detector. The fringe visibility $%
\mathcal{V}=\mathcal{V}_{0}\left\vert \mathrm{tr}_{D}\left( U\rho
_{D}\right) \right\vert $ is evidently smaller than the \textit{a priori}
fringe visibility, where $\rho _{D}$ is the initial state of the detector.
However, the measurement of an observable $\hat{W}$ of the detector can be
exploited to increase our knowledge about the which-path information.

A general strategy $\mathcal{S}$ of extracting the which-path information is
to split all the outcomes $W$ of measuring the observable $\hat{W}$ into two
disjoint sets $S$ and $\bar{S}$: if $W\in S$ then we guess that the quanton
takes path $|0\rangle $ and if $W\in \bar{S}$ then we guess that the quanton
takes path $|1\rangle $. By denoting
\begin{equation}
\eta _{S}=\sum_{W\in S}\langle W|\rho _{D}|W\rangle ,\;\eta
_{S}^{U}=\sum_{W\in S}\langle W|U\rho _{D}U^{\dagger }|W\rangle ,
\end{equation}%
and similarly for $\bar{S},$ the \textquotedblleft likelihood for guessing
the path right" is then given by $(1+\mathcal{D}_{\mathcal{S}})/2$, where
\begin{equation}
\mathcal{D}_{\mathcal{S}}\equiv 2w_{+}\eta _{S}+2w_{-}\eta _{\bar{S}}^{U}-1
\label{pd}
\end{equation}%
is the path distinguishability. By further choosing $\hat{W}$ to be such
that its eigenvectors $|W\rangle $ are also eigenvectors of $w_{+}\rho
_{D}-w_{-}U\rho _{D}U^{\dagger }$ and splitting the outcomes $W$ according to%
\begin{equation}
W\in \left\{
\begin{array}{ccc}
S, &  & \text{if }\langle W|w_{+}\rho _{D}|W\rangle >\langle W|w_{-}U\rho
_{D}U^{\dagger }|W\rangle \\
\bar{S}, &  & \text{if }\langle W|w_{+}\rho _{D}|W\rangle <\langle
W|w_{-}U\rho _{D}U^{\dagger }|W\rangle%
\end{array}%
\right. ,
\end{equation}%
the path distinguishability attains its maximum%
\begin{equation}
\mathcal{D}=\mathrm{tr}_{D}\left\vert w_{+}\rho _{D}-w_{-}U\rho
_{D}U^{\dagger }\right\vert .
\end{equation}%
By use of this mathematical expression of $\mathcal{D}$, Englert succeeded
in proving the Jaeger-Shimony-Vaidman-Englert inequality in Eq. (\ref{JSVE}).

In the following we shall identify two unsharp observables that are jointly
measured in the above experimental setup, or two noncommuting sharp
observables jointly measured in a nonideal way. The first unsharp observable
is described by the positive-operator-valued measure (POVM) $\mathcal{N}%
=\{N_{0},N_{1}\},$ where
\begin{equation}
N_{0}=\frac{1}{2}\left( I+\frac{\mathcal{V}}{\mathcal{V}_{0}}\sigma _{\delta
+\phi }\right) =I-N_{1}
\end{equation}%
with $\delta $ being the phase of $\mathrm{tr}_{D}(\rho _{D}U^{\dagger })$.
It is obviously a smeared version of the sharp observable $\sigma _{\delta
+\phi }.$ The probability of the quanton emerging from the output port $i$ ($%
i=0,1$) is $\mathrm{tr}_{Q}(\rho N_{i})$. Similar to Eq. (\ref{v0}), the
fringe visibility in this case can be written as
\begin{equation}
\mathcal{V}=\max_{\phi }(\langle N_{0}\rangle _{\rho }-\langle N_{1}\rangle
_{\rho }),
\end{equation}%
the maximum being attained when $\delta +\phi =\phi _{0}$.

The second unsharp observable, a smeared version of the sharp observable $%
\sigma _{x}$, is described by the POVM $\mathcal{M}=\{M_{0},M_{1}\},$ where
\begin{equation}
M_{0}=\frac{\eta _{S}+\eta _{S}^{U}}{2}I+\frac{\eta _{S}-\eta _{S}^{U}}{2}%
\sigma _{x}=I-M_{1}.
\end{equation}%
The probability of finding the detector in a state belonging to $S$ (or $%
\bar{S}$) is $\mathrm{tr}_{Q}(\rho M_{0})$ [or $\mathrm{tr}_{Q}(\rho M_{1})$%
].

In fact, the setup implements a measurement of a bivariate four-outcome
observable $\{E_{ij}{,\;i,j=0,1}\}$ where%
\begin{equation}
E_{i0}=\sum_{W\in S}E_{iW},\quad E_{i1}=\sum_{W\in \bar{S}}E_{iW},
\end{equation}%
together with%
\begin{eqnarray*}
E_{iW} &=&\mathrm{tr}_{D}[(I\otimes \rho _{D})\mathcal{U}^{\dagger
}(|i\rangle \langle i|\otimes |W\rangle \langle W|)\mathcal{U}], \\
&& \\
\mathcal{U} &=&(H\otimes I)U_{QD}(\Phi H\otimes I).
\end{eqnarray*}%
The $E_{ij}$'s satisfy the following two identities%
\begin{equation}
\sum_{j=0,1}E_{ij}=N_{i},\quad \sum_{i=0,1}E_{ij}=M_{j},  \label{mar}
\end{equation}%
so that recording the result $i$ is a measurement of the observable $%
\mathcal{N}$, while recording the result $j$ is a measurement of the
observable $\mathcal{M}$. In general, two unsharp observables $\mathcal{N}$
and $\mathcal{M}$ are jointly measurable if and only if there exists a
single experimental setup measuring a bivariate \textquotedblleft joint
observable" $\{E_{ij}\}$ whose marginals are $\mathcal{N}$ and $\mathcal{M}$
\cite{jm,jm1,Busch,BS}.

\section{Necessary and sufficient condition for joint measurability}

\label{sec:condition}

The unsharp observables of interest here are of the following kinds%
\begin{eqnarray}
N_{0} &=&\frac{I}{2}+\boldsymbol{n}\cdot \boldsymbol{\sigma }=I-N_{1},
\notag \\
M_{0} &=&m_{0}I+\boldsymbol{m}\cdot \boldsymbol{\sigma }=I-M_{1}  \label{mn}
\end{eqnarray}%
with ${\boldsymbol{n}\cdot \boldsymbol{m}}=0.$ Now, what is the criterion
for such a pair of unsharp observables to be jointly measurable? Only in the
special case where $m_{0}=1/2$ did Busch offer the answer to this question
\cite{Busch}. But in the case of a general $m_{0}$, this is an open question
so far. Here, we establish the following theorem:

\textit{Theorem 1. }Two unsharp observables $\mathcal{N}$ and $\mathcal{M}$
as given in Eq. (\ref{mn}) are jointly measurable if and only if
\begin{equation}
\sqrt{m_{0}^{2}-m^{2}}+\sqrt{(1-m_{0})^{2}-m^{2}}\geq 2n,  \label{thm1}
\end{equation}%
where $m=|{\boldsymbol{m}}|$ and $n=|{\boldsymbol{n}}|.$

\textit{Proof.} The most general forms of ${E_{ij}}$'s that take the $N_{i}$%
's and the $M_{j}$'s as marginals [i.e., that satisfy Eq. (\ref{mar})] read
as
\begin{equation*}
E_{ij}=x_{ij}I+\boldsymbol{y}_{ij}\cdot \boldsymbol{\sigma }
\end{equation*}%
with
\begin{eqnarray*}
x_{ij} &=&\tfrac{1}{4}+\tfrac{1}{4}(-)^{j}(2m_{0}-1)+(-)^{i+j}\tfrac{1}{2}x,
\\
&& \\
\boldsymbol{y}_{ij} &=&\tfrac{1}{2}\left[ (-)^{j}\boldsymbol{m}+(-)^{i}%
\boldsymbol{n}+(-)^{i+j}\boldsymbol{y}\right] ,
\end{eqnarray*}%
where $x$ is a real number and $\boldsymbol{y}$ is a vector in the
three-dimensional Euclidean space $\mathbb{R}^{3}$. The positivity of the $%
E_{ij}$'s entails the conditions $x_{ij}\geq \left\vert \boldsymbol{y}%
_{ij}\right\vert $ for all $i,j$. Therefore the joint measurability of $%
\mathcal{N}$ and $\mathcal{M}$ is equivalent to the existence of $x$ and $%
\boldsymbol{y}$ such that
\begin{eqnarray}
\left\vert \boldsymbol{m}+\boldsymbol{n}+\boldsymbol{y}\right\vert &\leq
&m_{0}+x,\quad \left\vert \boldsymbol{m}-\boldsymbol{n}+\boldsymbol{y}%
\right\vert \leq 1-m_{0}-x,  \notag \\
\left\vert \boldsymbol{m}-\boldsymbol{n}-\boldsymbol{y}\right\vert &\leq
&m_{0}-x,\quad \left\vert \boldsymbol{m}+\boldsymbol{n}-\boldsymbol{y}%
\right\vert \leq 1-m_{0}+x.  \notag \\
&&  \label{ineq}
\end{eqnarray}

Sufficiency. If the inequality Eq. (\ref{thm1}) holds then the choice $x=0$
and $\boldsymbol{y}=y\boldsymbol{n}/n$ with
\begin{equation}
y=\min \Big\{\sqrt{m_{0}^{2}-m^{2}}-n,n+\sqrt{(1-m_{0})^{2}-m^{2}}\Big\}
\end{equation}%
will make all four inequalities in Eq. (\ref{ineq}) hold. Hence a joint
observable can be explicitly constructed and the two unsharp observables $%
\mathcal{N}$ and $\mathcal{M}$ are joint measurable.

Necessity. Let us denote by $\mathcal{P}_{\boldsymbol{mn}}$ the plane
spanned by $\boldsymbol{m}$ and $\boldsymbol{n}$ in $\mathbb{R}^{3}$ (see
Fig. 2). A point $Y$ in $\mathbb{R}^{3}$ corresponds to a vector $%
\boldsymbol{y}\ $whose initial point is the original point $O$ and whose end
point is $Y,$ so when we say \textquotedblleft a point $\boldsymbol{y}$" we
mean the corresponding vector. The four points $A,$ $B,$ $C,$ and $D$
correspond to the vectors $-\boldsymbol{n}+\boldsymbol{m},$ $\boldsymbol{m}+%
\boldsymbol{n},$ $\boldsymbol{-m-n},$ and $\boldsymbol{n}-\boldsymbol{m},$
respectively. The four points $P,$ $Q,$ $S,$ and $T$ denote midpoints of the
line segments $AC,$ $BD,$ $AB,$ and $CD$, respectively. The first inequality
in Eq. (\ref{ineq}) means that the distance between the points $\boldsymbol{%
-m-n}$ and $\boldsymbol{y}$ is bounded by $m_{0}+x,$ and similarly for the
other three inequalities.

Suppose that the two unsharp observables $\mathcal{N}$ and $\mathcal{M}$ are
jointly measurable. Then there exist $x$ and $\boldsymbol{y}$ satisfying all
four inequalities in Eq. (\ref{ineq}). If $\boldsymbol{y}\notin \mathcal{P}_{%
\boldsymbol{mn}}$, then there must be a new point, e.g., the orthogonal
projector $\boldsymbol{y}^{\prime \prime }$ of $\boldsymbol{y}$ onto $%
\mathcal{P}_{\boldsymbol{mn}},$ that also satisfies the four inequalities in
Eq. (\ref{ineq}) with the same $x$. Further, if $\boldsymbol{y}^{\prime
\prime }=\overrightarrow{OY^{\prime \prime }},$ with $Y^{\prime \prime }$
being outside the rectangle $ABCD$ (see Fig. 2), then there must be a new
point inside the rectangle that also satisfies the four inequalities in Eq. (%
\ref{ineq}) with the same $x$. To see this, let $E$ denote the orthogonal
projection of $Y^{\prime \prime }$ onto the sideline $BD,$ and let $%
Y^{\prime }\in \square ABCD$ denote a point on the extension line of $%
Y^{\prime \prime }E$ satisfying $Y^{\prime }E\leq Y^{\prime \prime }E$. It
is evident that $AY^{\prime }<AY^{\prime \prime },$ $BY^{\prime }\leq
BY^{\prime \prime },$ $CY^{\prime }<CY^{\prime \prime },$ and $DY^{\prime
}\leq DY^{\prime \prime }$, so the vector $\boldsymbol{y}^{\prime }=%
\overrightarrow{OY^{\prime }}$ together with the same $x$ also satisfies the
four inequalities in Eq. (\ref{ineq}).

\begin{figure}[tbph]
\centering\includegraphics[width=75mm]{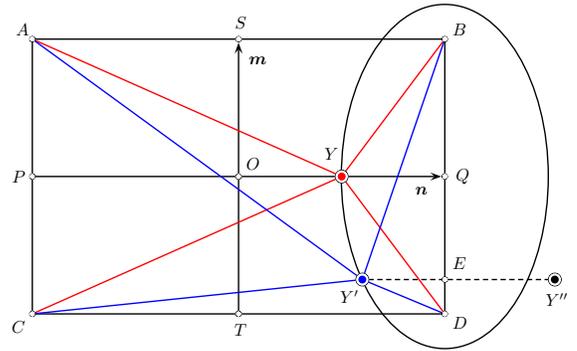}
\caption{Illustration of the fact that joint measurability of the two
unsharp observables $\mathcal{N}$ and $\mathcal{M}$ ensures the existence of
a point belonging to the line segment $PQ$ such that the corresponding
vector satisfies the four inequalities in Eq. (\protect\ref{ineq}) together
with $x=0$.}
\end{figure}

Then we will show that if $x$ and $\boldsymbol{y}^{\prime}\in \square ABCD$
satisfy the four inequalities in Eq. (\ref{ineq}), then there must be a
point belonging to the line segment $PQ$ which satisfies the four
inequalities in Eq. (\ref{ineq}) together with $x=0$. For the moment we will
consider the case $Y^{\prime }\in\square OQTD$ (see Fig. 2) and $x\geq 0$.
Other cases can be proved similarly. Let us denote by $\mathcal{E}%
_{BDY^{\prime}}$ the ellipse whose foci are $B$ and $D$ and which passes
through the point $Y^{\prime }$, and denote by $Y$ the point of intersection
of the ellipse with the straight line $PQ$. The point $Y$ may lie inside the
line segment $PQ$ or outside. For the moment we assume that it is inside the
line segment $PQ$. By assumption, we have%
\begin{eqnarray*}
AY^{\prime } &\leq &m_{0}-x,\quad BY^{\prime }\leq 1-m_{0}+x, \\
CY^{\prime } &\leq &m_{0}+x,\quad DY^{\prime }\leq 1-m_{0}-x.
\end{eqnarray*}
First, observe that%
\begin{equation*}
CY=AY\leq AY^{\prime }\leq m_{0}-x\leq m_{0}.
\end{equation*}
Then, let $\Delta $ be the difference between $BY^{\prime }$ and $BY$: $%
BY^{\prime }-BY=\Delta .$ Due to the property of an ellipse, we also have $%
DY-DY^{\prime }=\Delta .$ If $\Delta \geq x,$ then we have%
\begin{equation*}
DY=BY=BY^{\prime }-\Delta \leq 1-m_{0}+x-\Delta \leq 1-m_{0}.
\end{equation*}%
If $\Delta \leq x,$ then we have%
\begin{equation*}
BY=DY=DY^{\prime }+\Delta \leq 1-m_{0}-x+\Delta \leq 1-m_{0}.
\end{equation*}%
So $\overrightarrow{OY}$ is a vector that satisfies Eq. (\ref{ineq})
together with $x=0$.

If $Y$ is outside $PQ$, i.e., if it lies on the left of $P$, then following
the similar way one can show that $\overrightarrow{OP}$ satisfies the four
inequalities in Eq. (\ref{ineq}) together with $x=0$.

Similar arguments apply to the cases of $Y^{\prime }\notin \square OQTD$,
but in some cases we need to consider the ellipse $\mathcal{E}%
_{ACY^{\prime}} $ whose foci are $A$ and $C$ and which passes through the
point $Y^{\prime}$. We summarize that in the cases $\{x\geq
0,\;Y^{\prime}\in \square PQDC\}$ and $\{x\leq 0,\;Y^{\prime}\in \square
ABQP\}$ one should consider the ellipse $\mathcal{E}_{BDY^{\prime}},$ while
in the cases $\{x\leq 0,\;Y^{\prime}\in \square PQDC\}$ and $\{x\geq
0,\;Y^{\prime}\in \square ABQP\}$ one should consider the ellipse $\mathcal{E%
}_{ACY^{\prime}}.$

To sum up, a necessary condition for the two unsharp observables $\mathcal{N}
$ and $\mathcal{M}$ to be jointly measurable is the existence of a number $%
y\in \lbrack -n,n]$ such that%
\begin{equation}
m^{2}+(n+y)^{2}\leq m_{0}^{2},\quad m^{2}+(n-y)^{2}\leq (1-m_{0})^{2},
\end{equation}%
from which the inequality in Eq. (\ref{thm1}) immediately follows. \hfill$%
\square$

\section{Duality relation from joint measurability}

\label{sec:duality}

In Sec. \ref{sec:unsharp} we have seen that to each strategy $\mathcal{S}$
there correspond two unsharp observables $\mathcal{N}$ and $\mathcal{M}$ with%
\begin{eqnarray*}
\boldsymbol{n} &=&\frac{\mathcal{V}}{2\mathcal{V}_{0}}(0,\;-\sin {\phi _{0}}%
,\;\cos {\phi _{0}}), \\
m_{0} &=&\frac{\eta _{S}+\eta _{S}^{U}}{2}, \\
\boldsymbol{m} &=&\frac{\eta _{S}-\eta _{S}^{U}}{2}(1,\;0,\;0).
\end{eqnarray*}
Theorem 1 states that their joint measurability entails a bound on $\mathcal{%
V}/\mathcal{V}_{0}$:
\begin{equation}
\frac{\mathcal{V}}{\mathcal{V}_{0}}\leq \sqrt{\eta _{S}\eta _{S}^{U}}+\sqrt{%
\eta _{\bar{S}}\eta _{\bar{S}}^{U}}.  \label{thm1a}
\end{equation}%
In the following we will show that this joint measurability condition is
equivalent to a duality inequality.

From the definition Eq. (\ref{pd}) of the path distinguishability $\mathcal{D%
}_{\mathcal{S}}$ of a given strategy $\mathcal{S}$ and the identities $\eta
_{S}+\eta _{\bar{S}}=1,$ $\eta _{S}^{U}+\eta _{\bar{S}}^{U}=1,$ and $%
w_{+}+w_{-}=1,$ we obtain the following identity
\begin{equation*}
\mathcal{D}_{\mathcal{S}}^{2}+\left( \sqrt{\eta _{S}\eta _{S}^{U}}+\sqrt{%
\eta _{\bar{S}}\eta _{\bar{S}}^{U}}\right) ^{2}(1-\mathcal{P}^{2})=1-\gamma
_{\mathcal{S}}^{2},
\end{equation*}%
where
\begin{equation*}
\gamma _{\mathcal{S}}=2\Big|w_{+}\sqrt{\eta _{S}\eta _{\bar{S}}}-w_{-}\sqrt{%
\eta _{S}^{U}\eta _{\bar{S}}^{U}}\,\Big|.
\end{equation*}%
This identity together with the inequality in Eq. (\ref{thm1a}) leads us to
our main theorem:

\textit{Theorem 2. }When a general strategy $\mathcal{S}$ of extracting the
which-path information is adopted, the joint measurability condition of the
two unsharp observables $\mathcal{N}$ and $\mathcal{M}$ is \emph{equivalent}
to the following duality inequality characterizing the trade-off between the
fringe visibility and the path distinguishability:%
\begin{equation}
\mathcal{D}_{\mathcal{S}}^{2}+\frac{1-\mathcal{P}^{2}}{\mathcal{V}_{0}^{2}}%
\mathcal{V}^{2}\leq 1-\gamma _{\mathcal{S}}^{2}.  \label{main}
\end{equation}

Notice that this duality inequality is for a general strategy $\mathcal{S}$
and our derivation has not invoked the mathematical expression of the
optimal distinguishability $\mathcal{D}$, in sharp contrast to Englert's
derivation of the Jaeger-Shimony-Vaidman-Englert inequality. When we adopt
the optimal strategy $\mathcal{S}_{\mathrm{opt}}$ for extracting the
which-path information, i.e., when $\mathcal{D}_{\mathcal{S}}$ attains its
maximum $\mathcal{D}$, the Jaeger-Shimony-Vaidman-Englert inequality follows
immediately from our duality inequality. Although $\gamma _{\mathcal{S}_{%
\mathrm{opt}}}$ does vanish for a pure $\rho _{D}$ (which will be proved in
Appendix, Sec. \ref{sec:appendix a}), our duality inequality is more
stringent since $\gamma _{\mathcal{S}_{\mathrm{opt}}}$ does not generally
vanish. For example, if we take the detector to be a two-dimensional system
and let $w_{+}=(1+p)/2$ with $p$ being small then we have
\begin{equation}
\gamma _{\mathcal{S}_{\mathrm{opt}}}\approx \frac{2(1-\mathrm{tr}\rho
_{D}^{2})}{F(\rho _{D},U\rho _{D}U^{\dagger })}|p|+o(p^{2})  \label{gamma}
\end{equation}%
with $F$ being the quantum fidelity. It is clear that $\gamma _{\mathcal{S}_{%
\mathrm{opt}}}$ in this case does not vanish for any mixed $\rho _{D}$. See
Appendix, Sec. \ref{sec:appendix b} for a proof of Eq. (\ref{gamma}).

\section{Conclusions and discussions}

\label{sec:conclusion}

We have derived the necessary and sufficient condition for joint
measurability of two unsharp observables of the form in Eq. (\ref{mn}). This
is a substantial step towards solving the long-standing joint measurability
problem---given two unsharp observables, are they jointly measurable? In
fact, few such steps have ever been taken in the past two decades, since the
precursory work by Busch \cite{Busch}. We have also shown that our joint
measurability condition is equivalent to a duality relation between the
fringe visibility and the path distinguishability in Englert's setup, thus
establishing an intimate relationship between two different manifestations
of quantum complementarity.

Although our duality inequality is more stringent than the
Jaeger-Shimony-Vaidman-Englert inequality due to the quantity $\gamma _{%
\mathcal{S}}^{2}$ in Eq. (\ref{main}), we do not have a physical
interpretation of the quantity at present. In the past few years there have
appeared other duality inequalities different from the
Jaeger-Shimony-Vaidman-Englert kind. To take just one example, Jakob and
Bergou \cite{JakobBergou} established an intriguing inequality between the
local properties visibility and predictability and the nonlocal property
concurrence which is a quantitative entanglement measure. The relationship
between these duality inequalities and ours is yet unclear.

Another two open questions deserve further research. 1) The setup we
considered in this paper obviously simultaneously measures not merely
\textit{two} unsharp observables $\mathcal{N}$ and $\mathcal{M},$ but
\textit{three}: $\mathcal{N}$, $\mathcal{M}$, and $%
\{E_{00}+E_{11},E_{01}+E_{10}\}.$ One can expect that, although our present
formalism (attached to two observables) is enough for disclosing the
relationship between joint measurability and the duality relation, the
condition for the above three unsharp observables to be jointly measurable
will likely be tighter than that obtained in our present work, and so will
likely lead to a tighter duality inequality. However, a simple
single-inequality joint measurability condition for the three observables is
unavailable so far. 2) One might wonder whether we can follow our way to
derive a duality relation in a scenario more general than Englert's.
Although we have achieved a single-inequality joint measurability condition
for two general unsharp qubit observables \cite{Sixia}, and we have
succeeded in transforming the condition to a duality relation [Eq. (6) of
Ref. \cite{Sixia}] in \textit{some} more general scenarios, but we have
failed in doing so in \textit{the most general} scenarios.

\textit{Note added.} Recently, several works \cite%
{Sixia,buschnew,Heinosaari1,Heinosaari,Brougham,Busch08,stano} have appeared
on the topic of joint measurement of unsharp observables. For example, Busch
and Heinosaari \cite{buschnew} employed the notion of approximate joint
measurement (in some reasonable sense) to present some necessary (but not
sufficient) and some sufficient (but not necessary) conditions for two
unsharp qubit observables to be jointly measurable. Remarkably, the joint
measurability problem for two general unsharp qubit observables has been
solved by three independent groups \cite{Sixia,Busch08,stano}. Yu \textit{et
al.} \cite{Sixia} presented a single-inequality necessary and sufficient
condition for joint measurability and proved the equivalence between the
conditions formulated by the three groups.

\section*{ACKNOWLEDGEMENTS}

This work was supported by the NNSF of China, the CAS, the National
Fundamental Research Program (Grant No. 2006CB921900), and the Anhui
Provincial Natural Science Foundation (Grant No. 070412050).

\appendix

\section*{Appendix}

\renewcommand{\thesection}{A} \setcounter{equation}{0}

In this appendix, we present the proofs of some properties of $\gamma _{%
\mathcal{S}_{\mathrm{opt}}}.$

\subsection{Quantity $\protect\gamma _{\mathcal{S}_{\mathrm{opt}}}$ vanishes
when $\protect\rho _{D}$ is pure}

\label{sec:appendix a}

Since both $\rho _{D}$ and $\rho _{D}^{U}$ are pure, we can restrict
ourselves to the two-dimensional subspace spanned by these two pure states,
and we can denote them in terms of the Pauli operators acting on the
subspace:%
\begin{equation}
\rho _{D}=\frac{I+\boldsymbol{\alpha }\cdot \boldsymbol{\sigma }}{2},\quad
U\rho _{D}U^{\dagger }=\frac{I+\boldsymbol{\beta }\cdot \boldsymbol{\sigma }%
}{2},  \label{a1}
\end{equation}%
where $\boldsymbol{\alpha }$ and $\boldsymbol{\beta }$ are unit vectors in $%
\mathbb{R}^{3}.$ The optimal strategy for extracting the which-path
information is represented by the projective measurement $\{P_{\pm }\}$
\begin{equation}
P_{\pm }=\frac{I\pm \boldsymbol{s}\cdot \boldsymbol{\sigma }}{2},\quad
\boldsymbol{s}=\frac{w_{+}\boldsymbol{\alpha }-w_{-}\boldsymbol{\beta }}{%
|w_{+}\boldsymbol{\alpha }-w_{-}\boldsymbol{\beta }|}  \label{a2}
\end{equation}%
together with%
\begin{equation}
\eta _{S}=\frac{1+\boldsymbol{\alpha }\cdot \boldsymbol{s}}{2},\quad \eta
_{S}^{U}=\frac{1+\boldsymbol{\beta }\cdot \boldsymbol{s}}{2}.  \label{a3}
\end{equation}%
Straightforward algebra gives
\begin{equation}
w_{+}^{2}\eta _{S}\eta _{\bar{S}}-w_{-}^{2}\eta _{S}^{U}\eta _{\bar{S}}^{U}=%
\frac{1-\mathrm{tr}\rho _{D}^{2}}{2}p.  \label{a8}
\end{equation}%
Since $\rho _{D}$ is pure, we have $\mathrm{tr}\rho _{D}^{2}=1$ so that $%
w_{+}^{2}\eta _{S}\eta _{\bar{S}}-w_{-}^{2}\eta _{S}^{U}\eta _{\bar{S}%
}^{U}=0,$ i.e., $\gamma _{\mathcal{S}_{\mathrm{opt}}}=0.$

\subsection{Proof of Eq. (\protect\ref{gamma})}

\label{sec:appendix b}

Suppose states $\rho _{D}$ and $\rho _{D}^{U}$ of the detector (a
two-dimensional system) are of the form in Eq. (\ref{a1}) so that the
optimal strategy is given by Eqs. (\ref{a2}) and (\ref{a3}). For
convenience, let us denote $|\boldsymbol{\alpha }|^{2}=|\boldsymbol{\beta }%
|^{2}:=a$ and $\boldsymbol{\alpha }\cdot \boldsymbol{\beta }:=b.$ Since $%
\rho _{D}$ is a mixed state we have $a<1.$

The quantity $\gamma _{\mathcal{S}}$ can be rewritten as%
\begin{equation}
\gamma _{\mathcal{S}}=2\frac{\left\vert w_{+}^{2}\eta _{S}\eta _{\bar{S}%
}-w_{-}^{2}\eta _{S}^{U}\eta _{\bar{S}}^{U}\right\vert }{w_{+}\sqrt{\eta
_{S}\eta _{\bar{S}}}+w_{-}\sqrt{\eta _{S}^{U}\eta _{\bar{S}}^{U}}}.
\label{a4}
\end{equation}%
Using the fact that $p$ is small we obtain%
\begin{equation}
\frac{1}{w_{+}\sqrt{\eta _{S}\eta _{\bar{S}}}+w_{-}\sqrt{\eta _{S}^{U}\eta _{%
\bar{S}}^{U}}}=\frac{2}{\sqrt{1-\dfrac{a-b}{2}}}+o(p).  \label{a6}
\end{equation}%
On the other hand, the identity%
\begin{equation}
F(\rho _{D},U\rho _{D}U^{\dagger })=\sqrt{\mathrm{tr}(\rho _{D}U\rho
_{D}U^{\dagger })+2\det \rho _{D}}
\end{equation}%
together with%
\begin{equation}
\mathrm{tr}(\rho _{D}U\rho _{D}U^{\dagger })=\frac{1+b}{2},\quad \det \rho
_{D}=\frac{1-a}{4}
\end{equation}%
leads to%
\begin{equation}
F(\rho _{D},U\rho _{D}U^{\dagger })=\sqrt{1-\frac{a-b}{2}}.  \label{a5}
\end{equation}%
It follows from Eqs. (\ref{a6}) and (\ref{a5}) that%
\begin{equation}
\frac{1}{w_{+}\sqrt{\eta _{S}\eta _{\bar{S}}}+w_{-}\sqrt{\eta _{S}^{U}\eta _{%
\bar{S}}^{U}}}=\frac{2}{F(\rho _{D},U\rho _{D}U^{\dagger })}+o(p).
\label{a7}
\end{equation}%
Finally, inserting Eqs. (\ref{a8}) and (\ref{a7}) into Eq. (\ref{a4}) gives
Eq. (\ref{gamma}), thus completing the proof.


\begin{thebibliography}{99}
\bibitem{Bohr} N. Bohr, Naturwiss. \textbf{16}, 245 (1928); Nature (London)
\textbf{121}, 580 (1928).

\bibitem{Scully} M. O. Scully, B.-G. Englert, and H. Walther, Nature
(London) \textbf{351}, 111 (1991).

\bibitem{WZ} W. K. Wootters and W. H. Zurek, Phys. Rev. D \textbf{19}, 473
(1979).

\bibitem{GY} D. M. Greenberger and A. Yasin, Phys. Lett. A \textbf{128}, 391
(1988).

\bibitem{Mandel} L. Mandel, Opt. Lett. \textbf{16}, 1882 (1991).

\bibitem{Jaeger95} G. Jaeger, A. Shimony, and L. Vaidman, Phys. Rev. A
\textbf{51}, 54 (1995).

\bibitem{Englert} B.-G. Englert, Phys. Rev. Lett. \textbf{77}, 2154 (1996).

\bibitem{EnglertBergou} B.-G. Englert and J. A. Bergou, Opt. Commun. \textbf{%
179}, 337 (2000).

\bibitem{unsharp} Observables in terms of projection-valued measures are
called to be sharp, while observables in terms of positive-operator-valued
measures are called to be unsharp.

\bibitem{jm} H. Martens and W. M. de Muynck, Found. Phys. \textbf{20}, 255
(1990).

\bibitem{jm1} W. M. de Muynck, \textit{Foundations of Quantum Mechanics: An
Empiricist Approach} (Kluwer Academic Publishers, Dordrecht, 2002).

\bibitem{Busch} P. Busch, Phys. Rev. D \textbf{33}, 2253 (1986).

\bibitem{BS} P. Busch and C. Shilladay, Phys. Rep. \textbf{435}, 1 (2006).

\bibitem{JakobBergou} M. Jakob and J. A. Bergou, e-print
arXiv:quant-ph/0302075.

\bibitem{Sixia} S.\ Yu, N.-L. Liu, L. Li, and C. H. Oh, e-print
arXiv:0805.1538.

\bibitem{buschnew} P. Busch and T. Heinosaari, Quantum Inf. Comput. \textbf{8%
}, 797 (2008).

\bibitem{Heinosaari1} T. Heinosaari, P. Stano, and D. Reitzner, Int. J.
Quantum Inf. \textbf{6}, 975 (2008).

\bibitem{Heinosaari} T. Heinosaari, D. Reitzner, and P. Stano, Found. Phys.
\textbf{38}, 1133 (2008).

\bibitem{Brougham} T. Brougham, E. Andersson, and S. M Barnett, e-print
arXiv:0812.1474.

\bibitem{Busch08} P. Busch and H.-J. Schmidt, e-print arXiv:0802.4167.

\bibitem{stano} P. Stano, D. Reitzner, and T. Heinosaari, Phys. Rev. A
\textbf{78}, 012315 (2008).
\end{thebibliography}
\end{document}